\documentclass[english,preprint]{aastex}
\usepackage[T1]{fontenc}
\usepackage[latin9]{inputenc}
\usepackage{amssymb}

\makeatletter



\makeatother

\usepackage{babel}

\begin{document}

\title{A Four-Stokes-Parameter Spectral Line Polarimeter at the Caltech
Submillimeter Observatory}

\author{Talayeh Hezareh\altaffilmark{1}, \email{hezareh@astro.uwo.ca} Martin
Houde\altaffilmark{1}}

\altaffiltext{1}{The Department of Physics and Astronomy, The University of Western Ontario, London, Ontario, Canada N6A 3K7}

\begin{abstract}
We designed and built a new Four-Stokes-Parameter spectral line Polarimeter
(FSPPol) for the Caltech Submillimeter Observatory (CSO). The simple
design of FSPPol does not include any mirrors or optical components
to redirect or re-image the radiation beam and simply transmits the
beam to the receiver through its retarder plates. FSPPol is currently
optimized for observation in the $200-260$ GHz range and measures
all four Stokes parameters, $I$, $Q$, $U,$ and $V$. The very low
level of instrument polarization makes it possible to obtain reliable
measurements of the Goldreich-Kylafis effect in molecular spectral
lines. Accordingly, we measured a polarization fraction of a few percent
in the spectral line wings of $^{12}\mathrm{CO}$ $(J=2\rightarrow1)$
in Orion KL/IRc2, which is consistent with previous observations.
We also used FSPPol to study the Zeeman effect in the $N=2 \rightarrow 1$
transition of CN in DR21(OH) for the first time. At this point we
cannot report a Zeeman detection, but more observations are ongoing.
\end{abstract}

\keywords{Instrumentation: Polarimeters---ISM: Clouds---ISM: Magnetic Fields---Submillimeter}

\section{Introduction}

Astronomical polarimetry is a powerful tool for studying the characteristics
of the interstellar medium, from the large scale galactic magnetic
fields to the gravitational collapse of molecular cloud cores. The
understanding of the physical phenomena responsible for polarized
interstellar emission provides valuable information about the underlying
astrophysical processes, an important example being the formation
and evolution of stars. Since the 1990s, polarimetry techniques and
instruments have been developed for a wide range of wavelengths, from
optical (e.g., \citealt{Magalhaes,Wiktor}) to the submillimeter and
radio regimes (e.g., \citealt{Platt,Hildebrand 1997,hiroko,greaves,Li,thum,Hildebrand 2009,houde 2009,heiles1,heiles2,heiles3}). 

One of the important applications of astronomical polarimetry lies
in the study of the effect of magnetic fields on the early stages
of star formation. Hence, it is important to characterize the interstellar
magnetic fields around star-forming regions. Indeed, it is possible
to obtain the orientation and strength of the magnetic fields observationally,
as they leave their signature on the emission of the interstellar
dust grains and molecules. For example, the dust continuum radiation
becomes linearly polarized in a magnetic field \citep{Hildebrand 1999}.
At submillimeter wavelengths this polarization is perpendicular to
the field lines and a polarimetry map of dust continuum emission will
therefore reveal the plane-of-the-sky orientation of the magnetic
field. Furthermore, in the presence of anisotropic radiation, magnetic
fields cause the emission from gas molecules to also be linearly polarized
by a few percent \citep{goldreich}. It is also possible to determine
whether this polarization is aligned parallel or perpendicular to
the plane-of-the-sky component of the ambient magnetic field \citep{deguchi}.
Finally, the only way to directly measure the strength of interstellar
magnetic fields is through Zeeman line-broadening measurements. More
precisely, the line-of-sight component of the field is obtained by
measuring the circular polarization in the emission of the Zeeman
components of molecular spectral lines (\citealt{Crutcher 93,crutcher 99}). 

We describe the Four-Stokes-Parameter spectral line Polarimeter (FSPPol)
we recently designed and successfully commissioned at the Caltech
Submillimeter Observatory (CSO) in November 2008 in $\S 2$. We present
linear polarization measurements of $^{12}\mathrm{CO}$ $(J=2\rightarrow1)$
in Orion KL/IRc2 in $\S 3$ and preliminary Zeeman observations of
the $N=2 \rightarrow 1$ transition of CN in DR21(OH) in $\S 4$.
We end with a summary in $\S 5$.

\section{Instrument Description}

\begin{figure}
\epsscale{1.0}\plotone{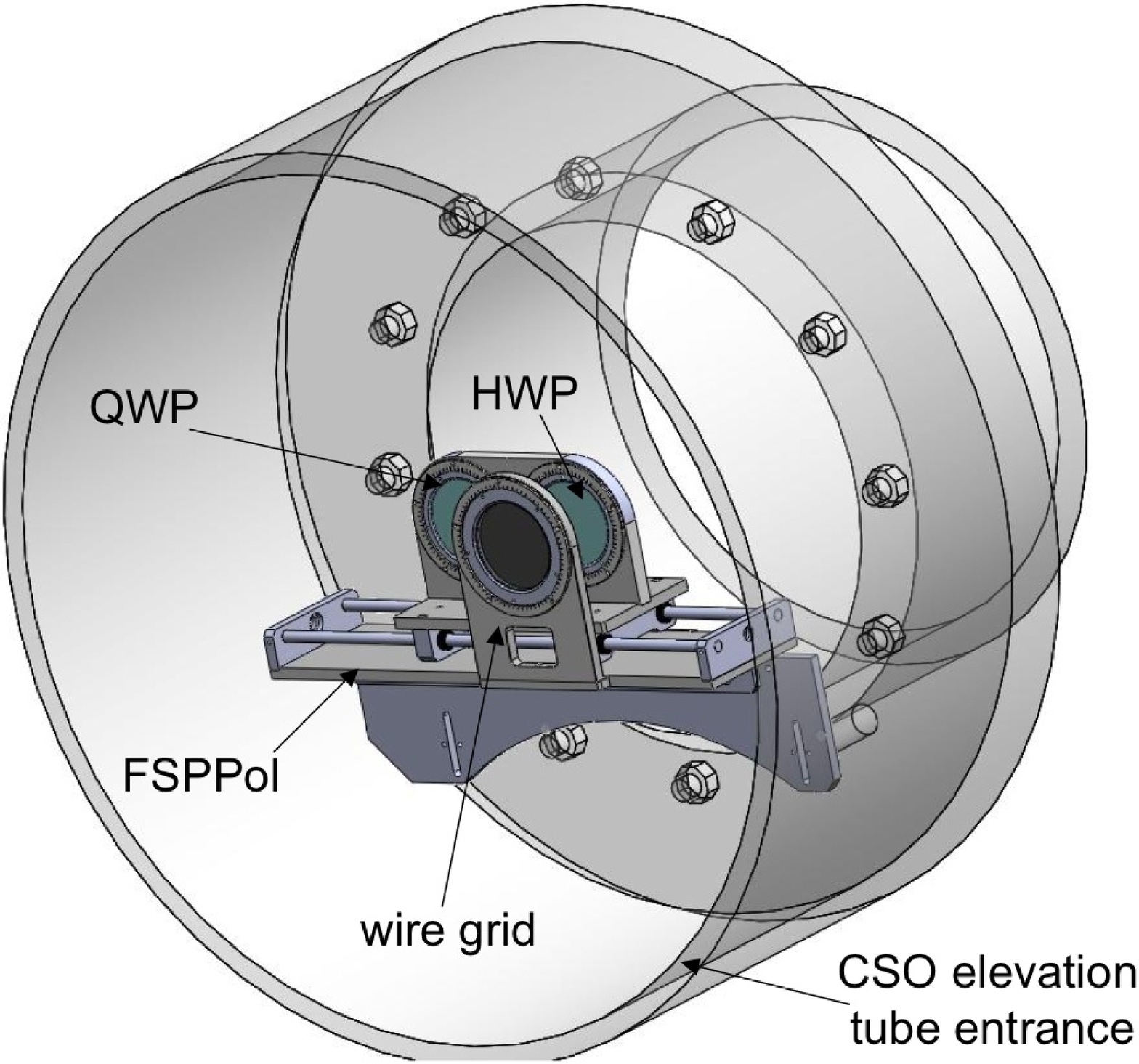}

\caption{\label{fig:polarimeter}The schematic diagram of FSPPol installed
in the elevation tube at the CSO. The polarizing grid, used for testing,
at the entrance of FSPPol is also shown.}

\end{figure}

\begin{figure}
\epsscale{1.0}\plotone{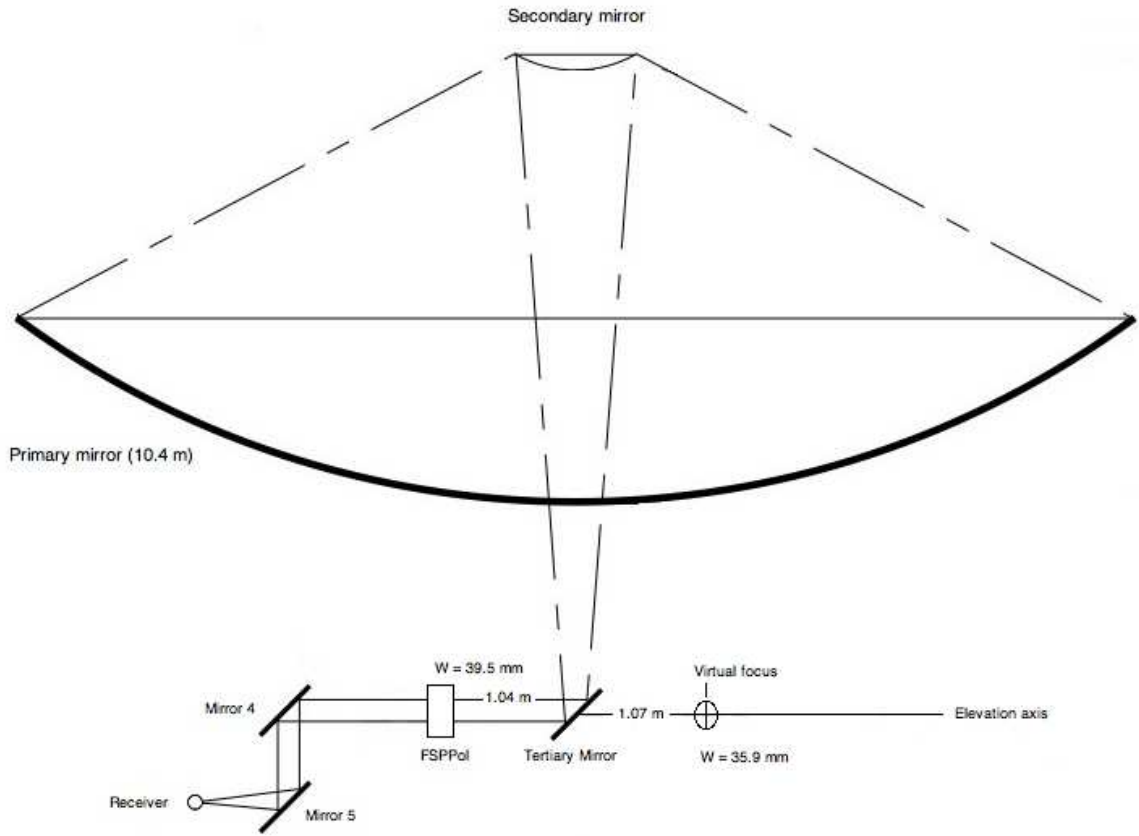}

\caption{\label{fig:cso}The schematic diagram of the CSO telescope, also showing
the position of FSPPol at 1.04 m from the tertiary mirror towards
Mirror 4. The telescope has an alt-azimuth mount and the receiver
is located at one of the two Nasmyth foci. The virtual focus of the
tertiary mirror is located 1.07 m behind the tertiary, and the beam
waists (W) at the virtual focus and at FSPPol are 35.9 mm and 39.5
mm, respectively. }

\end{figure}

The design of FSPPol is based on single beam polarimetry, as the current
heterodyne receivers at the CSO can measure only one linear polarization
state at one position at a time. The simple construction of the polarimeter
enabled us to mount the instrument inside the elevation tube that
traces much of the optical path from the tertiary mirror behind the
telescope dish to the Nasmyth focus where the $200-300$ GHz receiver
is located. A schematic diagram of FSPPol installed in the elevation
tube is shown in Figure \ref{fig:polarimeter}. The polarimeter is
mounted on a bracket attached to the walls of the tube and its location
with respect to the tertiary mirror and the receiver is depicted with
the schematic diagram of the CSO telescope in Figure \ref{fig:cso}.
The third focus of the $230$ GHz telescope beam is virtual and located
$1.07$ m behind the tertiary mirror and has a waist of $35.9$ mm.
We calculated the beam waist upon incidence on FSPPol to be $39.5$
mm. As illustrated in Figure \ref{fig:polarimeter}, FSPPol is comprised
of a half-wave plate (HWP) and a quarter-wave plate (QWP), each being
$100$ mm in diameter and optimized for observations at 226 GHz. The
wave plates were manufactured by Meller Optics, Inc.%
\footnote{120 Corliss Street, Providence, RI 02904 USA%
} and subsequently anti-reflection coated by QMC Instruments Ltd.%
\footnote{Cardiff University, School of Physics and Astronomy, The Parade, Cardiff
CF24 3AA UK%
}. More detailed specifications of the wave plates are listed in Table
\ref{ta:hwp-qwp}. These plates are installed in rotating rings mounted
side by side on an aluminum translational stage that moves across
the elevation tube, placing either wave plate in the path of the signal
reflected from the tertiary mirror. The anti-reflection coating on
the plates is efficient in eliminating standing waves along the optical
path. Nevertheless, the mounting of the translational stage was adjusted
in such a manner that the telescope beam impinged on the HWP or QWP
at an incidence that is slightly off from normal (i.e., at most a
few degrees) to avoid potential standing waves. The rotational and
translational motions of these wave plates are precisely monitored
and controlled by the instrument controlling software from the observatory's
control room. The system temperature increases by about $8\%$ when
the HWP is placed in the path of the beam, and by about $6\%$ with
the QWP in the beam. 

We aligned the optical axes of the wave plates in relation with the
polarization axis of the receiver by measuring the intensity from
a cold load, located at the entrance of FSPPol, while rotating the
wave plates over a wide range of angles. We previously placed a wire
grid with its polarizing state parallel to the receiver's axis between
the wave plates and the cold load to maintain a specific incoming
polarization state for testing purposes; this grid was removed for
astronomical observations. It is straightforward to show that the
location of a minimum in measured intensity with this set-up corresponds
to an angle where either the fast or slow axis of a plate (HWP or
QWP) is aligned with the receiver axis. This allows for a precise
orientation of the HWP and QWP with the receiver axis (also see $\S 3$).
It should also be noted that the slow axis is marked by the manufacturer
on each plate, which we used as a further consistency check. A small
circular cold load was also used to position the polarimeter within
the elevation tube such that the telescope beam was centered on the
surface of the HWP and QWP.


To test the instrument's accuracy, we observed the linear polarization
in the $J=2 \rightarrow 1$ transition of $^{12}$CO towards the Orion
KL/IRc2 high mass star forming region. After successful linear polarization
tests, we proceeded to perform Zeeman observations on the $N=2 \rightarrow 1$
transition of CN in DR21(OH) for the first time. The details of these
observations are explained in the following sections.

\section{Linear Polarization Measurements}

We observed $^{12}\mathrm{CO}\,\,\left(J=2\rightarrow1\right)$ at
$230.54$ GHz in three different locations in Orion KL/IRc2 coincident
with previous polarization observations by \citet{girart}, in order
to assess the performance of FSPPol. These locations, defined in the
equatorial coordinate system, were offset $(20\arcsec,20\arcsec)$,
$(20\arcsec,-20\arcsec)$ and $(-20\arcsec,-20\arcsec)$ from the
center of IRc2 ($\alpha=05^{\mathrm{h}}35^{\mathrm{m}}14.50^{\mathrm{s}}$
and $\delta=-05^{\circ}22^{'}30.4^{''}$, J2000.0). The observations
were performed between September $27^{\mathrm{th}}$ and October $6^{\mathrm{th}}$
2009 using the $200-300$ GHz receiver at the CSO, and the FFTS spectrometer
with a bandwidth of 500 MHz and a channel width of 61 kHz that corresponds
to a velocity resolution of about 0.08 km s$^{-1}$. The telescope
efficiency was determined with scans on Jupiter and calculated to
be $\simeq65$ \% for a beam width of $\simeq32\arcsec$(FWHM). During
these observations the typical optical depth at 225 GHz, as obtained
with the CSO radiometer, was $\tau_{225}\thickapprox0.06$ and the
typical system temperature with FSPPol in use was $T_{sys}\thickapprox370$
K. 

\begin{figure}
\epsscale{1.0}\plotone{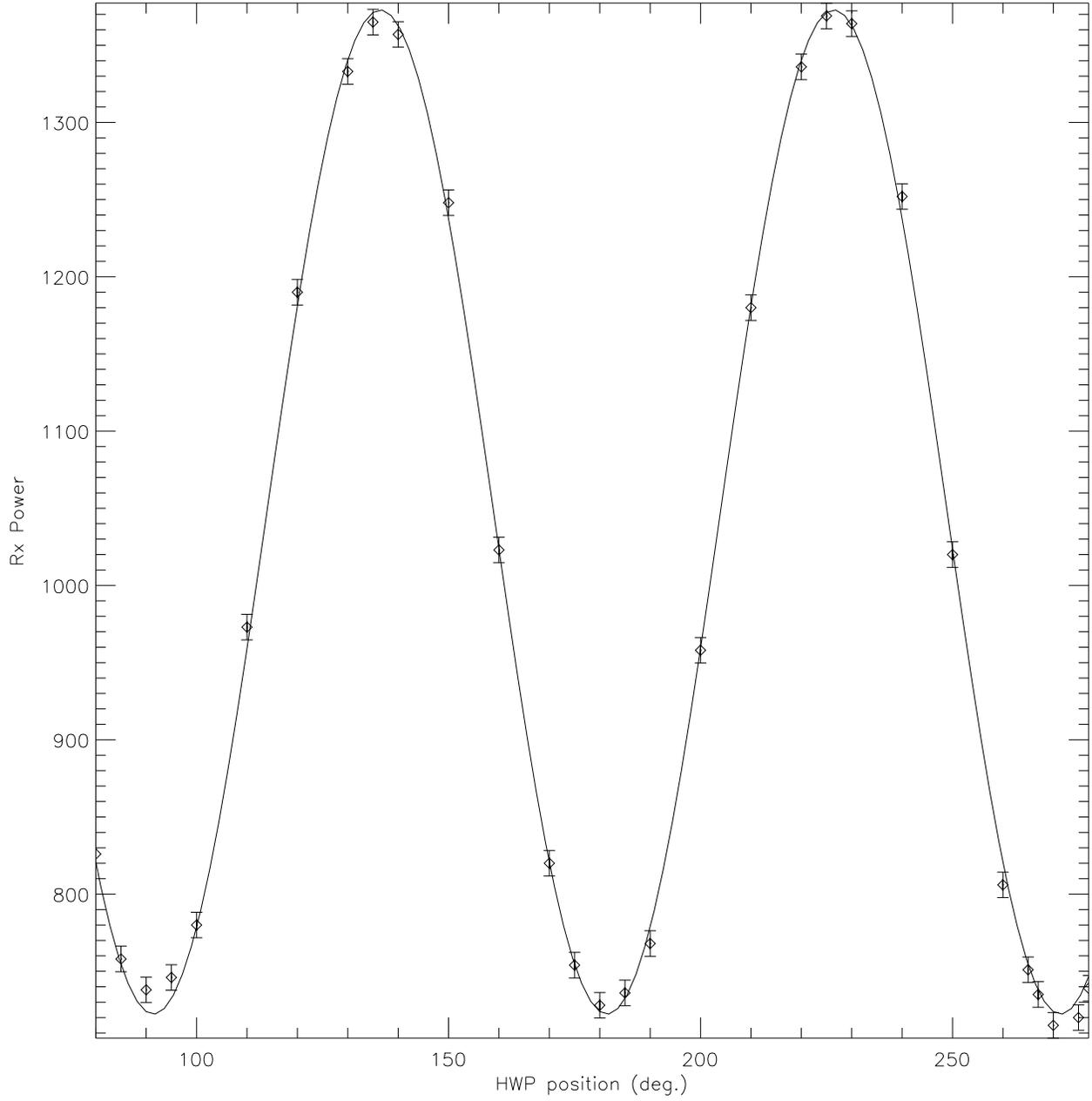}

\caption{\label{fig:hwp}The graph of the variation of the power measured from
a cold load with the CSO receiver against different orientations of
the HWP with respect to the receiver polarization axis. The $\mathrm{R}_{\mathrm{x}}$
power is in arbitrary units.}

\end{figure}

Although the refractive indices for the ordinary and extraordinary
rays vary significantly with frequency, the birefringence of crystal
Quartz appears to be relatively unchanged over a wide band \citep{afsar}
even though there is some uncertainty in the literature over its value
(0.048 for the frequencies we are concerned with; \citealp{marrone}).
However, since the HWP was operated at approximately 230.5 GHz, i.e.,
more than 4 GHz away from its design frequency, we characterized its
performance at the aforementioned frequency by measuring the power
from a cold load at different orientations of the HWP slow axis with
respect to the polarization axis of the facility receiver at the CSO.
Figure \ref{fig:hwp} shows the plot of the power at the receiver
against different angles between the HWP and receiver axis. The power
is measured in arbitrary units and the angles are displayed in degrees.
A polarizing grid, with its polarization state parallel to the receiver
axis, was placed between the cold load and the HWP to maintain a specific
incoming polarization state. As seen in the figure the HWP response
is very good at that frequency, although the power at successive minima
is observed to slightly vary \citep{savini}. Our model for the fit
to the data in Figure \ref{fig:hwp} is a cosine function with the
form

\begin{equation}
I(\theta)=\frac{(\mathrm{CL}+\mathrm{HL})}{2}+\frac{(\mathrm{CL}-\mathrm{HL})}{2}\mathrm{cos}(4(\theta-\delta)),\label{eq:hwp_fit}\end{equation}
 where $I(\theta)$ is the power measured by the receiver at angle
$\theta$ and $\delta$ determines the offset of the HWP slow axis
from the orientation defining the polarization state of the receiver.
CL is the power from the cold load and HL is the power due to emission
from the receiver that reflects back on the wires of the polarizing
grid; The HL signal incident on the HWP is therefore polarized perpendicular
to the CL signal. It should also be noted that there is an equal contribution
to CL and HL from receiver noise, although we cannot quantify it with
these measurements alone. The fit values for CL, HL and $\delta$
are $722.2\pm2.1$, $1373.0\pm2.9$ and $1.45^{\circ}\pm0.11^{\circ}$,
respectively. 

The polarization efficiency of FSPPol for linear polarization measurements
was previously determined in November 2008 by taking scans on Saturn.
We placed a wire grid at the entrance of FSPPol with its polarization
state parallel to the receiver's axis to force a precise incoming
polarization state, and observed Saturn for HWP angles $\theta=0^{\circ}$,
$90^{\circ}$, $45^{\circ}$ and $135^{\circ}$, where $\theta$ is
defined as the angle between the HWP slow axis and the receiver polarization
axis. We performed a one-minute (on-source) integration for each measurement,
with the system temperature being calibrated before every scan and
the signal from Saturn being integrated over the whole available spectral
bandwidth. Assuming that the wire grid is perfectly polarizing and
that Saturn is unpolarized, we obtained a polarization efficiency
of $\simeq99\%$. These measurements also allowed us to estimate the
instrument polarization to be on the order of $\simeq0.3\%$. In view
of this high polarization efficiency, we did not correct our Orion
KL/IRc2 data for the efficiency and instrumental effects.

\begin{figure}
\epsscale{1.0}\plotone{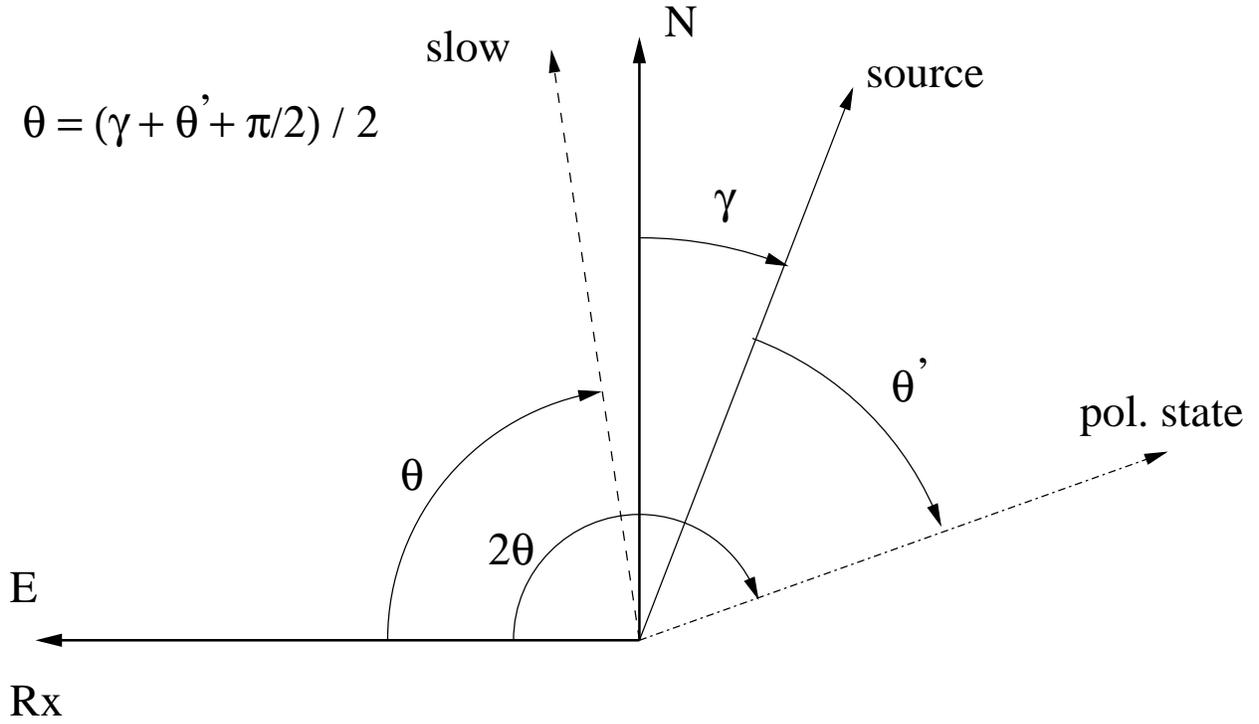}

\caption{\label{fig:coordinates}A schematic diagram for our linear polarization
measurements. The equatorial coordinate system shows North and East
on the sky. The receiver's polarization axis ($\mathrm{R}_{x}$) projected
on the sky in the equatorial system, lies along the East$-$West sky
axis, $\theta$ is the angle between $\mathrm{R}_{x}$ and the slow
axis of the HWP (denoted by {}``slow''), $\gamma$ is the parallactic
angle of the source, and $\theta^{'}$ the angle at which we want
to measure the polarization state in the frame of the source. Accordingly,
we have $\theta=(\gamma+\theta'+90^{\circ})/2$. }

\end{figure}

The polarimeter and the receiver are mounted in such a way that they
rotate in elevation with the telescope and the polarization axis of
the receiver on the sky, which is precisely oriented east-west (parallel
to the horizon) when FSPPol is not used, is preserved regardless of
the pointing of the telescope since the CSO telescope has an alt-azimuth
mount. Our aim for linear polarization measurements was to obtain
the Stokes $Q$ and $U$ parameters in the reference frame of the
sky, i.e., in the equatorial coordinate system, as well as the Stokes
$I$. Therefore, the rotation of the object's orientation on the sky,
or in other words the changes in its parallactic angle with time had
to be considered. For this purpose, we obtained one-minute (on-source)
long intensity measurements for different orientations of the HWP
slow axis relative to the receiver polarization axis $\theta=(\gamma+\theta^{'}+90^{\circ})/2$,
where $\gamma$ is the parallactic angle defining the orientation
of the object on the sky and $\theta^{'}$ is the angle at which we
seek to measure the linear polarization state in the reference frame
of the source (i.e., the frame that is rotated by $\gamma$ in the
equatorial system; see Figure \ref{fig:coordinates}). These measurements
are denoted by $I_{\theta^{'}}$. The instrument controlling software
continuously obtained the updated value of the parallactic angle from
the observatory's antenna computer during the on-source integration,
and compared it to the initial value at the beginning of the integration.
Once the change in $\gamma$ exceeded a predetermined threshold (i.e.,
$1^{\circ}$), the software commanded the antenna computer to stop
the integration, rotated the HWP by the updated angle using the new
value for $\gamma$, and the integration resumed. Our observations
were performed in cycles of four measurements for $\theta'=0^{\circ}$,
$90^{\circ}$, $45^{\circ}$ and $135^{\circ}$. Similar to the observations
on Saturn, we took one minute long scans for each measurement, and
calibrated the system temperature before every scan. We adopted this
conservative observing plan to minimize the effect of pointing or
calibration errors in the polarization data. The telescope pointing
was verified hourly with typical \textquotedbl{}five points\textquotedbl{}
integrations on CRL865 ($\alpha=06^{\mathrm{h}}{03}^{\mathrm{m}}{59.8}^{\mathrm{s}}$
and $\delta=07^{\circ}25^{'}51.4^{''}$, J2000.0) as our reference
star. 

\begin{figure}
\begin{center}  
\includegraphics[height=2.9in]{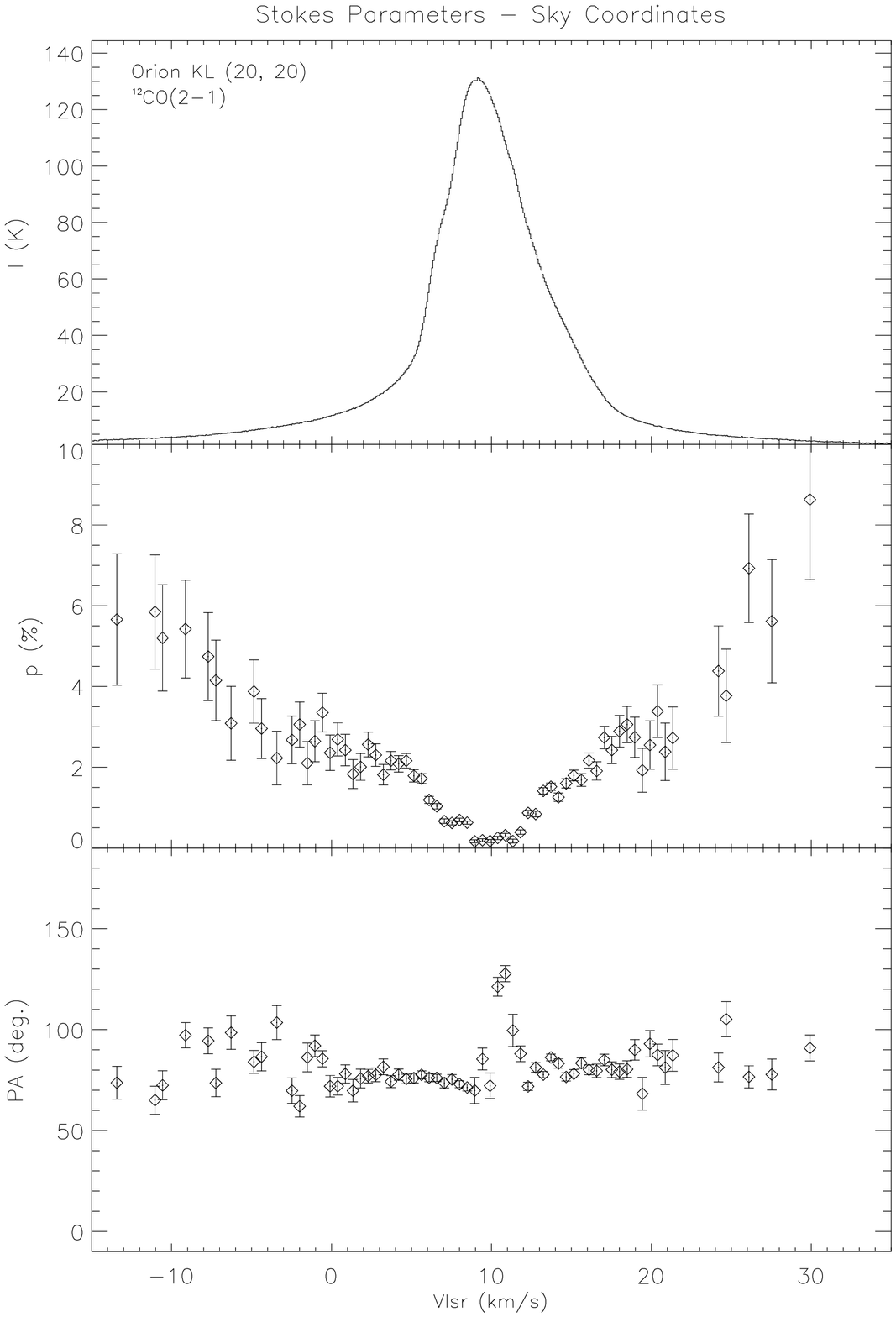}  
\includegraphics[height=2.9in]{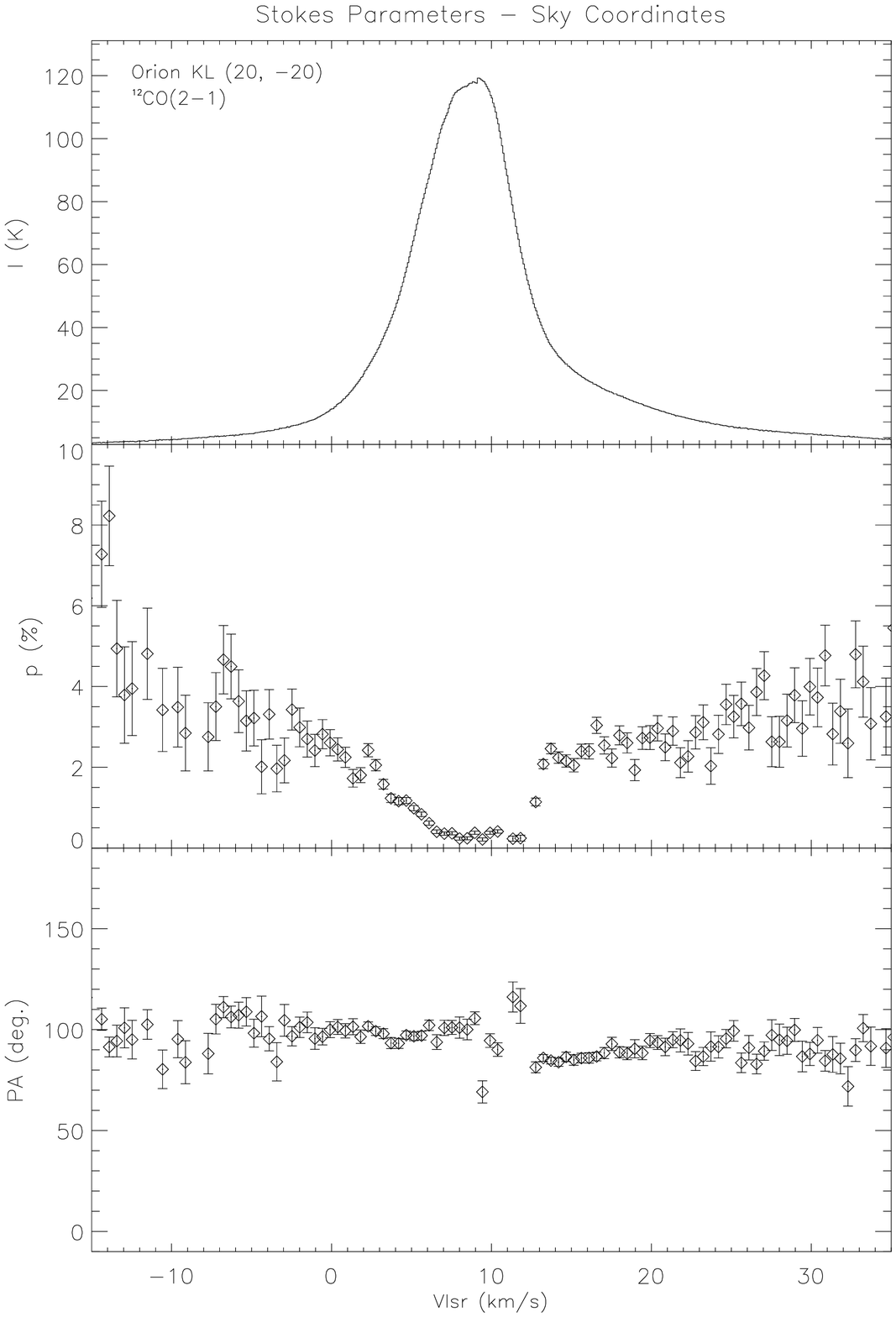}  
\includegraphics[height=2.9in]{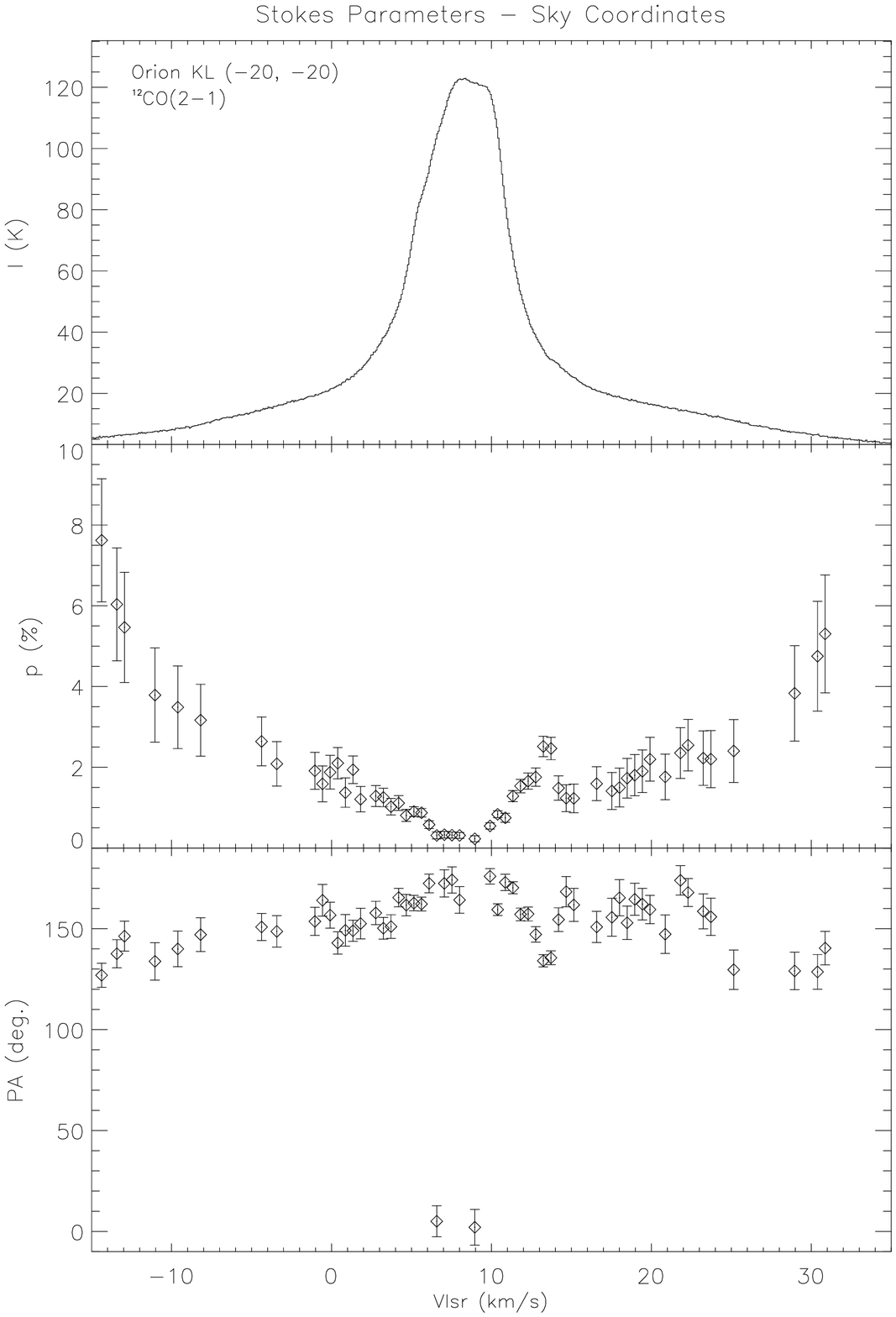}    
\end{center}  
\caption{Linear polarization measurements of $^{12}$CO ($J=2 \rightarrow 1$) in Orion KL/IRc2; ($top$) Stokes $I$, corrected for the telescope beam efficiency; ($middle$) The polarization level $p$ and ($bottom$)   angles PA across the spectral line.} 
\label{fig:1} 
\end{figure}

With the aforementioned definition for $I_{\theta^{'}}$ we have 

\begin{eqnarray}
I & = & \frac{\left(I_{0^{\circ}}+I_{90^{\circ}}+I_{45^{\circ}}+I_{135^{\circ}}\right)}{2}\nonumber \\
Q & = & I_{0^{\circ}}-I_{90^{\circ}}\label{eq:stokes}\\
U & = & I_{135^{\circ}}-I_{45^{\circ}},\nonumber \end{eqnarray}

\noindent and 

\begin{eqnarray}
p & = & \frac{\sqrt{Q^{2}+U^{2}}}{I}\label{eq:p-chi}\\
\mathrm{PA} & = & \frac{1}{2}\arctan\left(\frac{U}{Q}\right)\nonumber \end{eqnarray}

\noindent for the polarization fraction and angle (measured from
north increasing eastwards), respectively. The results from our measurements
are shown in Figure \ref{fig:1}, where the top graphs show the Stokes
$I$ spectra (corrected for the beam efficiency), while the middle
and bottom graphs are for the corresponding polarization fractions
and angles, respectively. For $p$ and $\mathrm{PA}$ the data were
binned using six adjacent velocity channels and only values for which
$p\ge3\sigma_{p}$ are plotted, where $\sigma_{p}$ is the uncertainty
in the polarization fraction. 

Table \ref{ta:co} displays the average values for $p$ and $\mathrm{PA}$
in the blue and red wings of the CO spectra, as well as the small
polarization fraction detected at the center of the lines. These values
were obtained after calculating the average of the Stokes $I$, $Q$
and $U$ across the stated velocity ranges. The small polarization
level in the center of the lines is probably the contribution from
the instrumental polarization, as it is expected that the linear polarization
due to the Goldreich-Kylafis effect will be greatly reduced where
the optical depth is high \citep{goldreich}. The values of $\mathrm{PA}$
are generally uniform across the spectral lines except for the center
of the lines, where the relative contribution from the instrumental
polarization is significant.

One potential problem in the linear polarization measurements is the
presence of polarized sidelobes that introduce false polarization
signals in the data \citep{frobrich}. This is more likely to happen
when the telescope is pointed towards the edge of an extended source
for the purpose of polarimetry measurements in low intensity regions.
This way, it is possible that strong emission from the core of the
source falls on these polarized sidelobes and contaminates the data.
We have yet to determine the potential contribution of polarized sidelobes
to data obtained with FSPPol at the CSO, we cannot therefore comment
on their significance for our results on Orion KL. We hope to do so
during a future observing run.

Our polarization results are in general consistent with the findings
of \citet{girart} (see their Figure 1). The polarization fraction
that we calculate in the spectral line wings of $^{12}$CO is in good
agreement with their results, and the difference between their values
for polarization angles and ours (i.e., $10^{\circ}-20^{\circ}$)
could solely be the result of different telescope beam sizes. This
could also explain the slight differences in line profiles and intensities
between their Stokes $I$ spectra and ours. For example the secondary
peak at $\approx12$ km $\mathrm{s}^{-1}$ in the offset positions
obtained by \citet{girart} is not resolved in our spectral data.

\section{Zeeman Measurements of CN ($N=2 \rightarrow 1$)}

As mentioned earlier, measurement of the Zeeman effect in interstellar
molecular spectral lines is the only direct way to obtain the strength
of magnetic fields in molecular clouds. The Zeeman line-broadening
in an interstellar spectral line profile is directly proportional
to the strength of the ambient magnetic field, and can be studied
by observing the signature of circular polarization in the line profile.
To this date, Zeeman detections have been reported in the spectral
lines of a few species, namely \ion{H}{1}, OH, $\mathrm{H_{2}O}$,
$\mathrm{CH_{3}\mathrm{OH}}$ and CN \citep{sarma09,troland,falgarone,sarma01,crutcher 99,plante}.
The high critical density of CN makes this molecule suitable for studying
magnetic fields in dense regions. Additionally, the rotational transition
lines of CN contain several hyperfine components that have different
Zeeman splitting coefficients. This makes it possible to distinguish
between the true Zeeman effect and instrumental effects that produce
artificial circular polarization in the data \citep{crutcher 96}.
Although the Zeeman effect in the $N=1 \rightarrow 0$ transition
of CN has previously been detected \citep{falgarone,crutcher 99},
to the best of our knowledge it has never been attempted at $N=2 \rightarrow 1$.
The higher critical density of the latter will make it possible to
probe denser regions in molecular clouds and better establish observationally
how the magnetic fi{}eld strength scales with density. 


There are nine strong hyperfine components in the $N=2 \rightarrow 1$
transition of CN. The frequencies, Zeeman coefficients and relative
intensities of these lines are displayed in Table \ref{ta:hf_z}.
Similar to the $N=1 \rightarrow 0$ transition previously observed
in DR21(OH), the CN ($N=2 \rightarrow 1$) lines are double peaked,
suggesting the existence of two velocity components \citep{crutcher 99}.
These two velocity components may be arising from different regions,
and may exhibit different field strengths. It is possible to observe
the nine hyperfine components simultaneously and fit the observed
circular polarization data to the following expression for both velocity
components \citep{crutcher 99}

\begin{equation}
V_{i}=a(I_{i1}+I_{i2})+b_{1}(\frac{dI_{i1}}{d\nu})+b_{2}(\frac{dI_{i2}}{d\nu})+c_{1}(Z_{i}\frac{dI_{i1}}{d\nu})+c_{2}(Z_{i}\frac{dI_{i2}}{d\nu}),\label{eq:vfit}\end{equation}
 where $V_{i}$ is the total circular polarization for hyperfine line
$i$. In the above equation, the different sources of polarization
are expressed in separate terms. The first term represents the polarization
contribution from the error in intensity calibrations in the two polarization
modes, $a$, that adds a small image of the intensity spectrum, $I_{i1}+I_{i2}$,
to the $V_{i}$ spectra. The second and third terms are polarization
contributions from the beam squint effect represented by $b_{1}$
and $b_{2}$, that is caused when the right and left handed polarization
measurements do not probe exactly the same region, for which a non-zero
velocity gradient exists in its vicinity \citep{Crutcher 93}. This
effect is brought about by pointing errors of the telescope, and also
in some cases by mechanical deformations in the antenna. The last
two terms are the true Zeeman signals for each velocity component
of the spectrum with $c_{1}$ and $c_{2}$ representing the strength
of the line-of-sight component of the magnetic fields $B_{\mathrm{los}1}/2$
and $B_{\mathrm{los}2}/2$, respectively, for each velocity component,
and $Z_{i}$ being the Zeeman coefficient for hyperfine line $i$.
This way, the real Zeeman effect is separated from instrumental effects
by determining the above fitting parameters.

\subsection{Observations}

We observed CN ($N=2 \rightarrow 1$) at 226.8 GHz in DR21(OH) ($\alpha=20^{\mathrm{h}}39^{\mathrm{m}}01^{\mathrm{s}}$
and $\delta=42^{\circ}22^{'}37.7^{''}$, J2000.0) during July $8^{\mathrm{th}}$
to $13^{\mathrm{th}}$ and September $27^{\mathrm{th}}$ to October
$6^{\mathrm{th}}$ 2009, using the FFTS spectrometer with bandwidth
of 500 MHz and channel resolution of $61$ kHz. The telescope efficiency
was determined with scans on Saturn and Jupiter and calculated to
be $\simeq65$ \% for a beam width of $\simeq32\arcsec$(FWHM). We
verified the telescope pointing on an approximately hourly rate by
performing $^{12}$CO ($J=2 \rightarrow 1$) scans on $\chi$ Cygni
($\alpha=19^{\mathrm{h}}50^{\mathrm{m}}33.8^{\mathrm{s}}$ and $\delta=32^{\circ}{54}^{'}53.2^{''}$,
J2000.0) as our reference star. Our first observing session suffered
from issues with an unstable receiver and mediocre skies. The average
system temperature at that time was about $500$ K and the typical
$\tau_{225}$ was $\approx0.12$. The circular polarization observations
were performed by taking scans of $\mathrm{CN}$ ($N=2 \rightarrow 1$)
in DR21(OH) with the QWP slow axis rotated by $+45^{\circ}$ and $-45^{\circ}$
with respect to the receiver's polarization axis. With our instrument
set-up, our one minute on-source integrations at $+45^{\circ}$ project
the right-handed circular polarization ($I_{RCP}$) emission (on the
sky) on the receiver's axis, while similar integrations at $-45^{\circ}$
probe the corresponding left-handed circular polarization ($I_{LCP}$)
emission. The Stokes $I$ and $V$ are obtained from $I=I_{LCP}+I_{RCP}$
and $V=I_{LCP}-I_{RCP}$. As before, this observing strategy was chosen
in order to mitigate any potential calibration or pointing errors.
The system temperature was calibrated before every scan. We have so
far obtained a total on-source integration time of 553 minutes, with
an average system temperature of 450 K.

\begin{figure}
\epsscale{1.0}\plotone{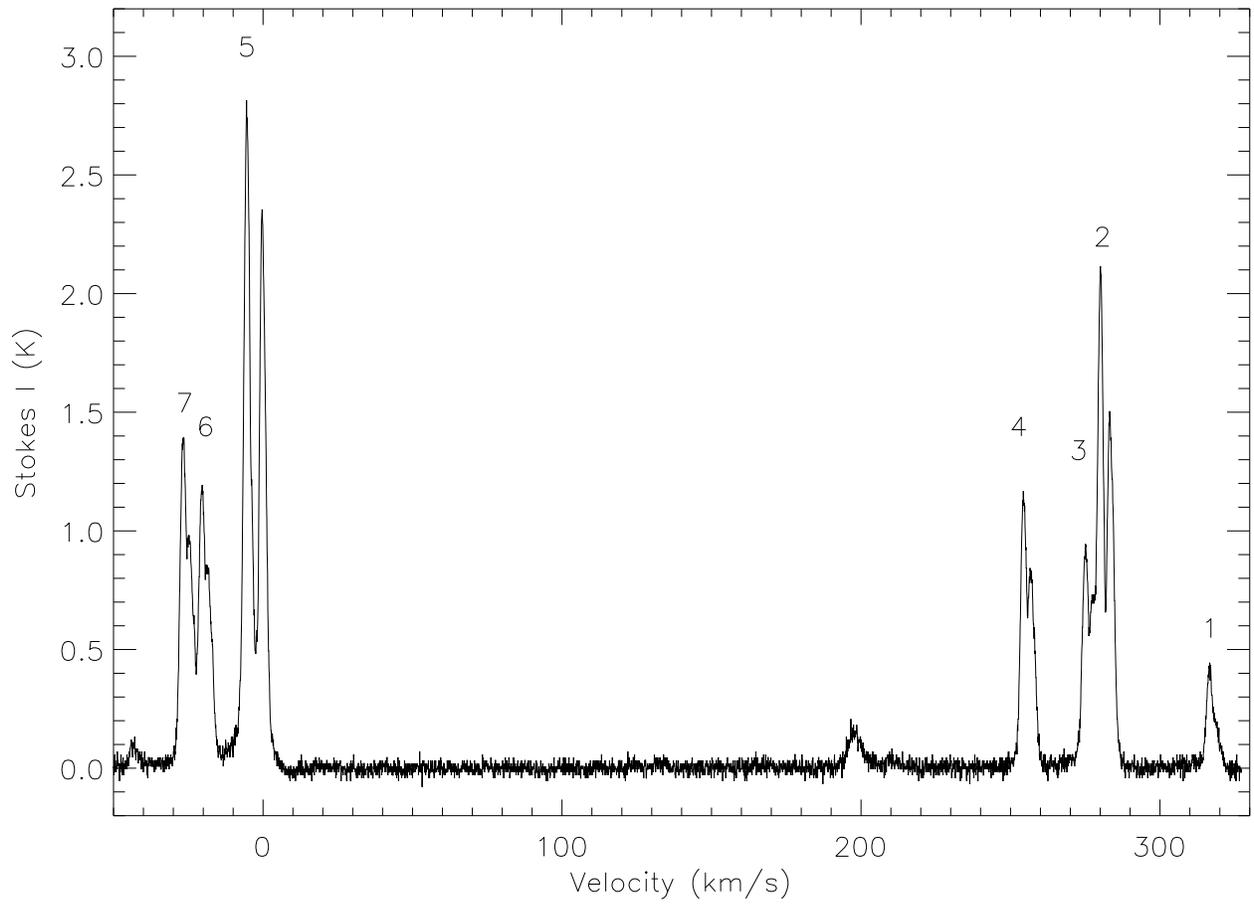}

\caption{\label{fig:spectra}Stokes $I$ spectrum of the nine strong hyperfine
components of CN ($N=2 \rightarrow 1$) towards DR21(OH). Line 5 is
actually three blended lines that appear as one. }

\end{figure}

Figure \ref{fig:spectra} shows the Stokes $I$ spectrum for the CN
($N=2 \rightarrow 1$) hyperfine lines that are labeled according
to the order given in Table \ref{ta:hf_z} with the line temperatures
corrected for the telescope beam efficiency. There are two further
hyperfine lines at $\simeq-45$ km s$^{-1}$ and $\simeq200$ km s$^{-1}$
that are weak relative to the labeled lines and therefore not included
in the study. Observing the $N=2 \rightarrow 1$ transition of CN
brings more complications compared to the $N=1 \rightarrow 0$ transition.
For example, there are three lines in Table \ref{ta:hf_z} that are
labeled 5 because they are heavily blended and appear as a single
broad line in the observed spectrum, as seen in Figure \ref{fig:spectra}.
Furthermore, due to the double peaked feature of the CN lines, the
other hyperfine lines with small frequency separations (i.e., lines
2 and 3 and lines 6 and 7) are blended together as well, although
to a lesser level. 

An inspection of the ratio of line temperatures reveals that our CN
lines are not consistent with the LTE assumption. The spacing between
the two velocity components of every hyperfine line varies with the
line strength, and the rotational diagram for the $N=2 \rightarrow 1$
lines produced a negative excitation temperature, suggesting that
the lines are affected by self absorption. A more complete spectral
line analysis will be presented in a forth-coming paper.

\subsection{Preliminary Analysis \& Discussion}

The following is only a preliminary analysis on the data we have gathered
so far, as more data will be obtained in upcoming observing runs,
and a more thorough discussion will be presented then. 

Since the CN hyperfine lines in DR21(OH) are double peaked, we fitted
two to three Gaussian profiles to each line depending on its shape.
In the fitting procedure, the line widths of the Gaussian profiles
within each velocity component of a hyperfine line profile were kept
the same. For the three blended hyperfine components of line 5 (in
Figure \ref{fig:spectra}), we fitted one Gaussian profile to each
component and fixed their relative frequencies according to the information
given in Table \ref{ta:hf_z}. Each hyperfine component of the other
pairs of blended lines (lines 2 and 3 and lines 6 and 7) were treated
as single lines, such that Gaussian profiles incorporating both lines
were fitted to the blended line profiles. Once the fitting parameters
for all the Gaussian profiles were obtained, the blended lines could
thus be separated by subtracting the Gaussian fit of one line from
the spectrum of the pair. The Gaussian fits for the Stokes $I_{i}=I_{i1}+I_{i2}$
for both velocity components of every hyperfine line $i$, together
with their derivatives with respect to frequency and corresponding
Zeeman coefficients were simultaneously fitted to the Stokes $V$
spectrum using Equation (\ref{eq:vfit}). The resulting fitting parameters
were $a=-0.0015\pm0.0007$, $b_{1}=1.24\pm1.13$ kHz, $b_{2}=1.28\pm1.47$
kHz, $c_{1}=-0.5\pm1.3$ mG and $c_{2}=-2.5\pm2.0$ mG. Although we
have not obtained a Zeeman detection, the instrumental polarization
contribution is seen to be very small for $a$ and kept to reasonable
levels for $b_{1}$ and $b_{2}$ (i.e., of the same order as the expected
values for $c_{1}$ and $c_{2}$). 

\begin{figure}
\epsscale{1.0}\plotone{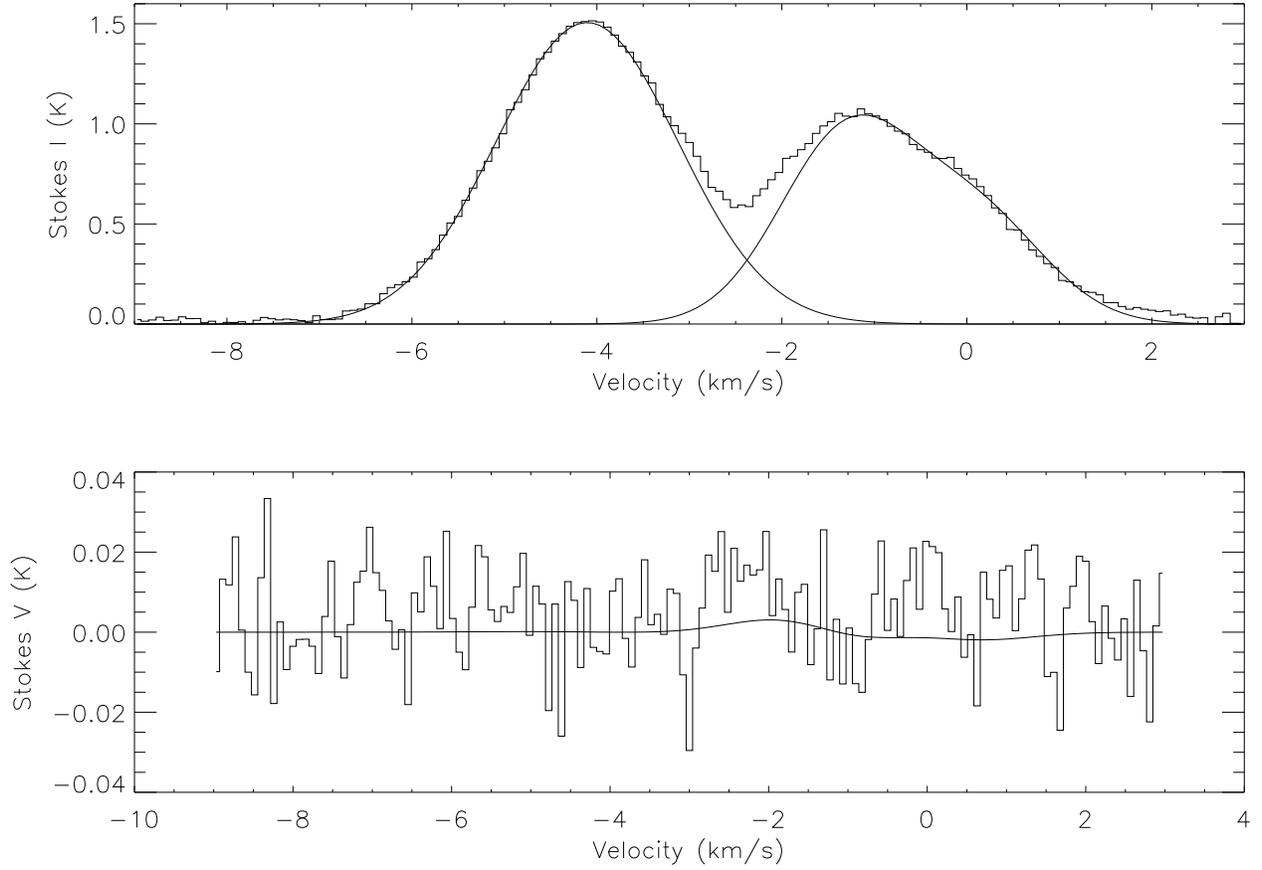}

\caption{\label{fig:spec_neg}($top$) The averaged Stokes $I$ spectrum for
the CN ($N=2 \rightarrow 1$) hyperfine lines 1, 2 \& 4 in DR21(OH)
that have negative Zeeman coefficients. The solid lines are the Gaussian
fits to the two velocity components of Stokes $I$ spectrum. ($bottom$)
The averaged Stokes $V$ spectrum for the same hyperfine lines. The
solid line is the fit for the Zeeman expression to the Stokes $V$
spectrum, with instrumental effects removed. }

\end{figure}

Since it is not easy to distinguish the Zeeman signal in the $V$
spectra of individual lines from the noise level, \citet{crutcher 99}
produced an averaged sum of the $N=1 \rightarrow 0$ hyperfine lines
with strong Zeeman coefficients to display the Zeeman fit for their
detection. In the case of the $N=2 \rightarrow 1$ transition, not
all hyperfine lines have the same sign for their Zeeman coefficients
and therefore it is only possible to average together the lines with
same coefficient signs. The top panel of Figure \ref{fig:spec_neg}
shows the weighted average Stokes $\overline{I}$ for lines 1, 2 and
4 that have negative Zeeman coefficients, with the weights being the
relative sensitivity to the magnetic field, i.e., $|\mathrm{Z_{\mathit{i}}\times R.I_{\mathit{i}}}|$
given in Table \ref{ta:hf_z}. All the hyperfine lines are centered
on the $V_{LSR}$ of the source, i.e., -3 km s$^{-1}$ and in order
to remove the contamination of line 3 from line 2, the Gaussian fit
for line 3 was subtracted from the blended spectrum. A similar procedure
was performed for the averaged $V$ spectrum of the aforementioned
lines, which is displayed in the lower panel of Figure \ref{fig:spec_neg}
with the instrumental effects removed. The bold line is the Zeeman
fit to the average $V$ data, expressed by $\overline{V}=\overline{Z}(C_{1}d\overline{I_{1}}/d\nu)+\overline{Z}(C_{2}d\overline{I_{2}}/d\nu)$,
where $\overline{Z}$ is the weighted average of the negative Zeeman
coefficients, with the weights being $|\mathrm{R.I_{\mathit{i}}}|$.
The fits for $C_{1}$ and $C_{2}$ are $C_{1}=-0.15\pm2.25$ mG and
$C_{2}=-3.57\pm3.22$ mG, which are consistent with the values obtained
for $c_{1}$ and $c_{2}$ from the simultaneous Zeeman fit to individual
$V$ spectra mentioned above.

The Zeeman Stokes $V$ signal due to a field of a few hundred $\mu$Gauss
is small (i.e., a line broadening of a few hundred Hz), and the noise
level in the spectra needs to be sufficiently low in order to obtain
at least a $2\sigma$ detection. We developed simulations for the
observed CN lines with different noise levels and magnetic field strengths,
to estimate the remaining integration time required to obtain a detection.
\citet{crutcher 99} obtained a value of $B_{\mathrm{los}}\simeq0.75$
mG for one of the velocity components of DR21(OH) by observing the
$N=1 \rightarrow 0$ transition of CN, which is associated with a
critical density $n_{c}\approx10^{5}\mathrm{cm^{-3}}$. Assuming that
the magnetic field strength varies with $n^{1/2}$, where $n$ is
the the average gas density of the observed region, we should expect
that the magnetic field we probe with our observations will be somewhat
stronger. Since $n_{c}\approx10^{6}\mathrm{cm^{-3}}$ at the $N=2 \rightarrow 1$
transition, $B_{\mathrm{los}}$ should be at most a few times stronger
than the value obtained with the $N=1 \rightarrow 0$ transition.
Our simulations show that for a 2 mG field, we need to have our noise
level down to $\simeq5$ mK to be able to get at least a $2\sigma$
detection. The RMS noise in our data is currently $\simeq16$ mK,
implying that we need more observations to obtain a credible detection.
Assuming an average system temperature of 400 K, with a bandwidth
of 61 kHz, we will require about 48 more hours of on-source integration
time. This time estimate is comparable with the observing time that
\citet{crutcher 99} spent on OMC1n and DR21OH to obtain a Zeeman
detection in the CN ($N=1 \rightarrow 0$) transition.

\section{Summary}

We recently designed and successfully commissioned a Four-Stokes-Parameter
spectral line Polarimeter (FSPPol) at the CSO in November 2008. The
simple design of FSPPol does not contain any mirrors or grids to redirect
or split the radiation beam, and the instrument is conveniently mounted
in the elevation tube between the tertiary mirror behind the telescope
dish and the Nasmyth focus where the heterodyne receiver is located.
FSPPol transmits the beam to the receiver through half-wave and quarter-wave
plates that are presently optimized for observations at 226 GHz.

We used FSPPol for linear and circular polarization measurements in
the spectral lines of interstellar molecules during the months of
July, September and October 2009. We measured a linear polarization
level of $\simeq1\%$ to $2.5\%$ due to the Goldreich-Kylafis effect
in the spectral line wings of $^{12}\mathrm{CO}$ $(J=2\rightarrow1)$
in Orion KL/IRc2, and our results are consistent with previous observations
\citep{girart}. We also started Zeeman observations on the $N=2 \rightarrow 1$
transition of CN in DR21(OH) for the first time. At this point we
have obtained about 10 hours of on-source integration time, and our
preliminary data analysis shows that although we have not detected
a Zeeman signal in the CN ($N=2 \rightarrow 1$) lines, the overall
contribution from the instrumental effects in the Stokes $V$ spectrum
is low. Further observations of CN ($N=2 \rightarrow 1$) are ongoing
for this source and other star forming regions.

\acknowledgements{The authors thank B. Dalrymple, S. H. Chen and  T. Officer for their
help in the development of FSPPoL, H. Shinnaga for helping with the
Zeeman calculations and M. Azimlu and S. Chitsazzadeh for assistance
during observations. M. H.'s research is funded through the NSERC
Discovery Grant, Canada Research Chair, Canada Foundation for Innovation,
Ontario Innovation Trust, and Western's Academic Development Fund
programs. The Caltech Submillimeter Observatory is funded through
the NSF grants AST 08-38261 and AST 05-40882 to the California Institute
of Technology. }

\clearpage

\clearpage

\begin{deluxetable}{ll}

\tabletypesize{\footnotesize}

\tablecaption{Properties of the wave plates \label{ta:hwp-qwp}}

\tablecolumns{2}

\tablewidth{0pt}

\tablehead{

\colhead{Property} & \colhead{}  \\

}

\startdata

Material  & single crystal Quartz         \\
Indices of refraction: &           \\
 Ordinary  &  $2.106 \pm 0.006 \tablenotemark{a}$        \\
 Extraordinary  & $2.154 \pm 0.007$ \tablenotemark{a}       \\
Thickness  &  13.82 mm (HWP), 6.90 mm (QWP)     \\
Anti-reflection coating:  &           \\
 Material  &   High density Polypropylene (HDPP)     \\

\enddata
\tablenotetext{a}{Birch et al. 1994, IEEE: Transactions on Microwave Theory and Techniques, 42, 6, 956}
\end{deluxetable}

\clearpage

\begin{deluxetable}{cccccrr}

\tabletypesize{\footnotesize}

\tablecaption{Polarization measurements on CO $J=2 \rightarrow 1$ in Orion KL/IRc2 \label{ta:co}}


\tablecolumns{7}

\tablewidth{0pt}

\tablehead{

\colhead{} & \multicolumn{3}{c}{$p$ ($\%$)} & \colhead{} & \multicolumn{2}{c}{PA (deg)} \\

\cline{2-4} \cline{6-7} 

\colhead{Offsets (arcsec)\tablenotemark{a}} & \colhead{Blue\tablenotemark{b} wing} & \colhead{Line center} & \colhead{Red\tablenotemark{c} wing} & \colhead{} &

\colhead{Blue\tablenotemark{b} wing}& \colhead{Red\tablenotemark{c} wing}  

}

\startdata

(20$\arcsec$, 20$\arcsec$) & 2.1 $\pm$ 0.3 & 0.35 $\pm$ 0.04 & 

1.8 $\pm$ 0.2 & & 77.1 $\pm$ 4.2 & 81.2 $\pm$ 3.4   \\

(20$\arcsec$, -20$\arcsec$) & 1.6 $\pm$ 0.2 &  0.29 $\pm$ 0.04 & 

2.4 $\pm$ 0.2 & & 97.8 $\pm$ 3.3 & 88.1 $\pm$ 2.9   \\

(-20$\arcsec$, -20$\arcsec$) & 1.1 $\pm$ 0.3 &  0.29 $\pm$ 0.07 & 

1.5 $\pm$ 0.4 & & 154.8 $\pm$ 7.1 & 155.6 $\pm$ 8.1   \\

\enddata

\tablenotetext{a}{Offsets are with respect to IRc2 ($\alpha=05^{\mathrm{h}}35^{\mathrm{m}}14.5^{\mathrm{s}}$, $\delta = -05\arcdeg22\arcmin30.4\arcsec$; J2000.0).}
\tablenotetext{b}{$p$ and PA are averaged over the velocity range of  -4 to 6 km s$^{-1}$ (see Figure \ref{fig:1}).}
\tablenotetext{c}{$p$ and PA are averaged over the velocity range of  13 to 23 km s$^{-1}$.}
\end{deluxetable}

\clearpage

\begin{deluxetable}{cccccc}

\tabletypesize{\footnotesize}

\tablecaption{Hyperfine components of CN ($N=2 \rightarrow 1$)\tablenotemark{a} \label{ta:hf_z}}

\tablecolumns{6}

\tablewidth{0pt}

\tablehead{

\colhead{Line} & \colhead{$(N,J,F)\rightarrow(N',J',F')$} &\colhead{$\nu$ (MHz)} & \colhead{$Z$ (Hz/$\mu$G)} & \colhead{R.I\tablenotemark{b}}& \colhead{|$Z\times$R.I|\tablenotemark{c}}\\

}

\startdata

1 & (2,3/2,3/2)$\rightarrow$(1,1/2,3/2)  & 226632.19  & -0.72241 & 0.59259 & 0.42809   \\
2 & (2,3/2,5/2)$\rightarrow$(1,1/2,3/2)  & 226659.58  & -0.70995 & 2.0     & 1.41991     \\
3 & (2,3/2,1/2)$\rightarrow$(1,-1/2,1/2) & 226663.70  &  0.62277 & 0.59259 & 0.36905   \\
4 & (2,3/2,3/2)$\rightarrow$(1,1/2,1/2)  & 226679.38  & -1.18326 & 0.74074 & 0.87649   \\
5 & (2,5/2,5/2)$\rightarrow$(1,3/2,3/2)  & 226874.17  &  0.70995 & 2.016   & 1.43127   \\
5 & (2,5/2,7/2)$\rightarrow$(1,3/2,5/2)  & 226874.75  &  0.40035 & 3.200   & 1.28112   \\
5 & (2,5/2,3/2)$\rightarrow$(1,3/2,1/2)  & 226875.90  &  1.18326 & 1.200   & 1.41991   \\
6 & (2,5/2,3/2)$\rightarrow$(1,3/2,3/2)  & 226887.35  &  1.46973 & 0.3840  & 0.56438   \\
7 & (2,5/2,5/2)$\rightarrow$(1,3/2,5/2)  & 226892.12  &  1.05692 & 0.3840  & 0.40586   \\

\enddata
\tablenotetext{a}{Shinnaga et al. (2010), in preperation.}
\tablenotetext{b}{Relative intensity.}
\tablenotetext{c}{Relative sensitivity to $B$$_{\mathrm{los}}$.}

\end{deluxetable}

\end{document}